\documentclass[prb,reprint,a4paper,showpacs,citeautoscript,floatfix]{revtex4-1}
\pdfoutput=1

\usepackage{amsmath}
\usepackage[pdftex]{graphicx}
\usepackage{microtype}
\usepackage[charter]{mathdesign}
\usepackage{bm}\let\vec\bm

\usepackage[svgnames]{xcolor}
\usepackage[
	colorlinks=True,linkcolor=DarkRed,citecolor=ForestGreen,urlcolor=MediumBlue,
	pdfstartview=FitH,bookmarks=False,pdfpagemode=UseNone
]{hyperref}

\begin{document}

\title{First direct observation of the Van Hove singularity in the tunneling spectra of cuprates}

\author{A. Piriou}
\email{alexandre.piriou@unige.ch}
\author{N. Jenkins}
\author{C. Berthod}
\author{I. Maggio-Aprile}
\author{{\O}. Fischer}
\affiliation{DPMC-MaNEP, University of Geneva, 24 quai Ernest-Ansermet, 1211 Geneva 4, Switzerland}

\date{March 4, 2011}

\begin{abstract}

In two-dimensional lattices the electronic levels are unevenly spaced, and the density of states (DOS) displays a logarithmic divergence known as the Van Hove singularity (VHS). This is the case in particular for the layered cuprate superconductors. The scanning tunneling microscope (STM) probes the DOS, and is therefore the ideal tool to observe the VHS. No STM study of cuprate superconductors has reported such an observation so far giving rise to a debate about the possibility of observing directly the normal state DOS in the tunneling spectra. In this study, we show for the first time that the VHS is unambiguously observed in STM measurements performed on the cuprate Bi$_2$Sr$_2$CuO$_{6+\delta}$ (Bi-2201). Beside closing the debate, our analysis proves the presence of the pseudogap in the overdoped side of the phase diagram of Bi-2201 and discredits the scenario of the pseudogap phase crossing the superconducting dome.\\[0.5em]
\href{http://dx.doi.org/10.1038/ncomms1229}{DOI: 10.1038/ncomms1229}

\end{abstract}

\maketitle

\section*{Introduction}

The scanning tunneling spectroscopy (STS) is known to probe the density of states \cite{Tersoff-1983, Chen-1993}, and a logarithmic peak is therefore expected in the tunneling spectrum of materials having a VHS near the Fermi level.  However, there is a strong theoretical debate around the possibility to observe such features in tunneling spectra. The compelling absence of such a peak in the tunneling spectra of the cuprates has lead to claims that the normal-state DOS would in fact be canceled like just as in ideal planar tunnel junctions \cite{Harrison-1961, Yusof-1998}. An alternative is that the VHS peak is suppressed due to the interaction of electrons with collective excitations \cite{Eschrig-2000, Hoogenboom-2003b, Levy-2008}.

Among the cuprates, Bi-2201 has a combination of two properties required for an optimal STS investigation of the VHS: it is nearly two-dimensional, and has a low maximal $T_c$ of around 12$~$K giving access to the normal state at low temperature. Producing large and pure non cation-doped single crystals of Bi-2201 is difficult and consequently, this compound has attracted little attention. We succeeded in growing such crystals and tuning their macroscopic doping level \cite{Piriou-2010}, allowing doping dependent STS studies.

In this study, we demonstrate that the VHS is directly seen in the tunneling spectra of Bi-2201. We observe that all the spectra can be fitted by a model involving a VHS and a $d$-wave BCS gap. This suggests that the VHS shows up in the low-$T_c$ Bi-2201 due to a weak coupling to collective modes. Moreover, our measurement at a temperature above $T_c$ of a heavily overdoped Bi-2201 sample clearly exhibit a gap feature that we interpret as the signature of the pseudogap phase.

\section*{Results}

\subparagraph*{\hspace*{-\parindent}STM data for overdoped Bi-2201 samples.}

\begin{figure}[t]
\includegraphics[width=\columnwidth]{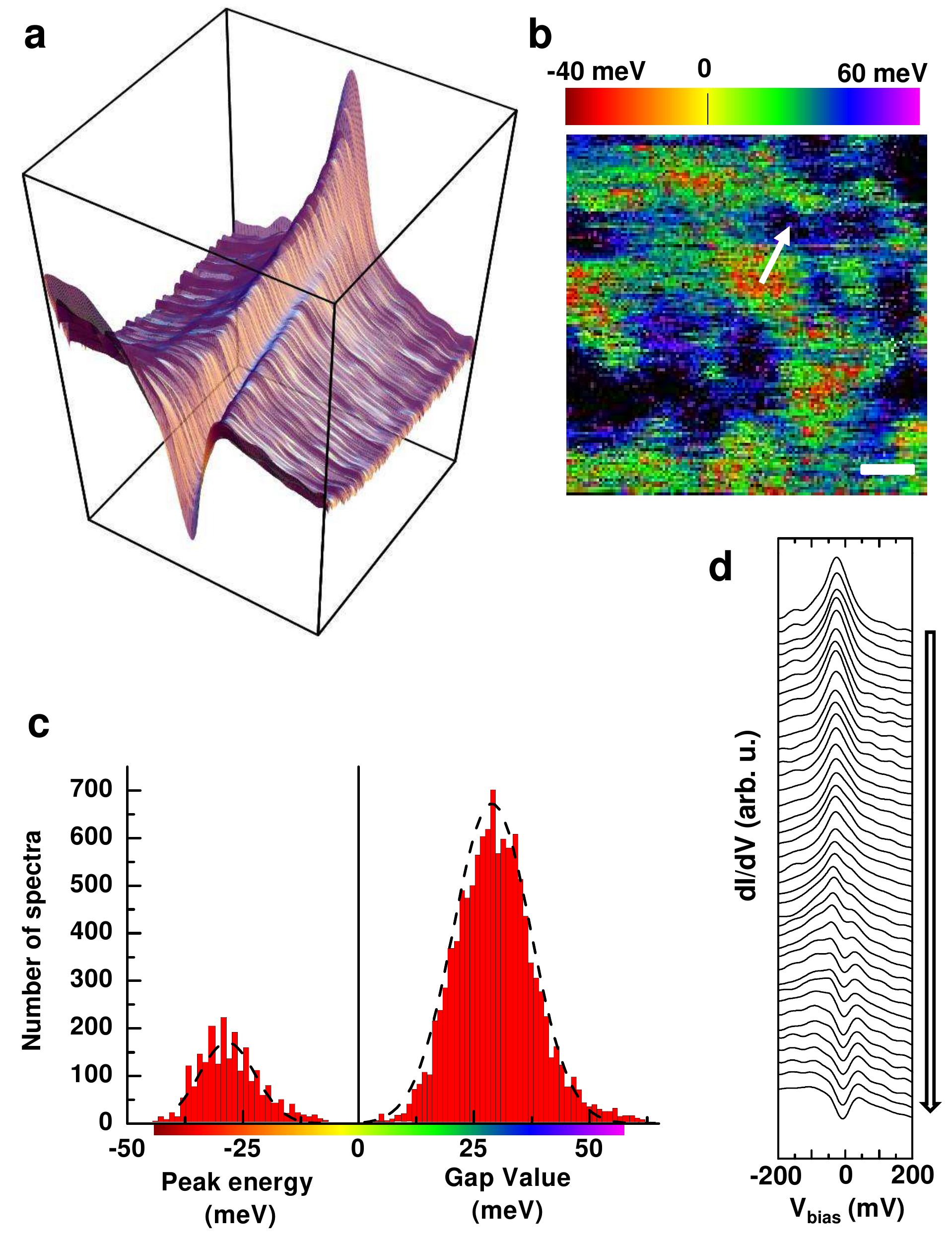}
\caption{\label{fig:fig1}
\textbf{Low-temperature (5~K) STS spectroscopy of as-grown Bi-2201 (AG11K).} (\textbf{a}) Three dimensional overview of the data: each curve is the average of 60 spectra sharing the same $E_p$ value ($E_p$ is the peak position for single-peak spectra, and half the peak-to-peak distance for two-peak spectra); the curves are sorted by increasing $E_p$. Note the smooth transition between completely different curves at extremal values of $E_p$. (\textbf{b}) $E_p$-map of a $14\times14$~nm$^2$ area. The white bar corresponds to 2~nm. The red and yellow regions correspond to negative $E_p$ (single-peak spectra) while in green, blue, and purple regions the spectra exhibit a gap of width $\Delta_p=E_p$. Red-yellow spots are surrounded by green regions where the smallest gap are measured ($\sim20$~meV). Black patches are composed of spectra ($\sim4000$ in total) in which no gap could be extracted due to a missing left peak (or sometimes no peak at all); these spectra  are not included in the histogram of panel \textbf{c}. (\textbf{c}) Distribution of $E_p$ values. The dashed lines are Gaussian fits. The average peak position and average gap values are $-28$ and $+29$~meV, respectively. (\textbf{d}) Spatial evolution of the spectra on the path depicted by the arrow in \textbf{b}. The path goes from a red region (top-most single-peak spectrum) to a black region where the left coherence peak is barely visible. Notice the strong similarity with the global evolution shown in \textbf{a}.
}
\end{figure}

\begin{figure}[t]
\includegraphics[width=0.7\columnwidth]{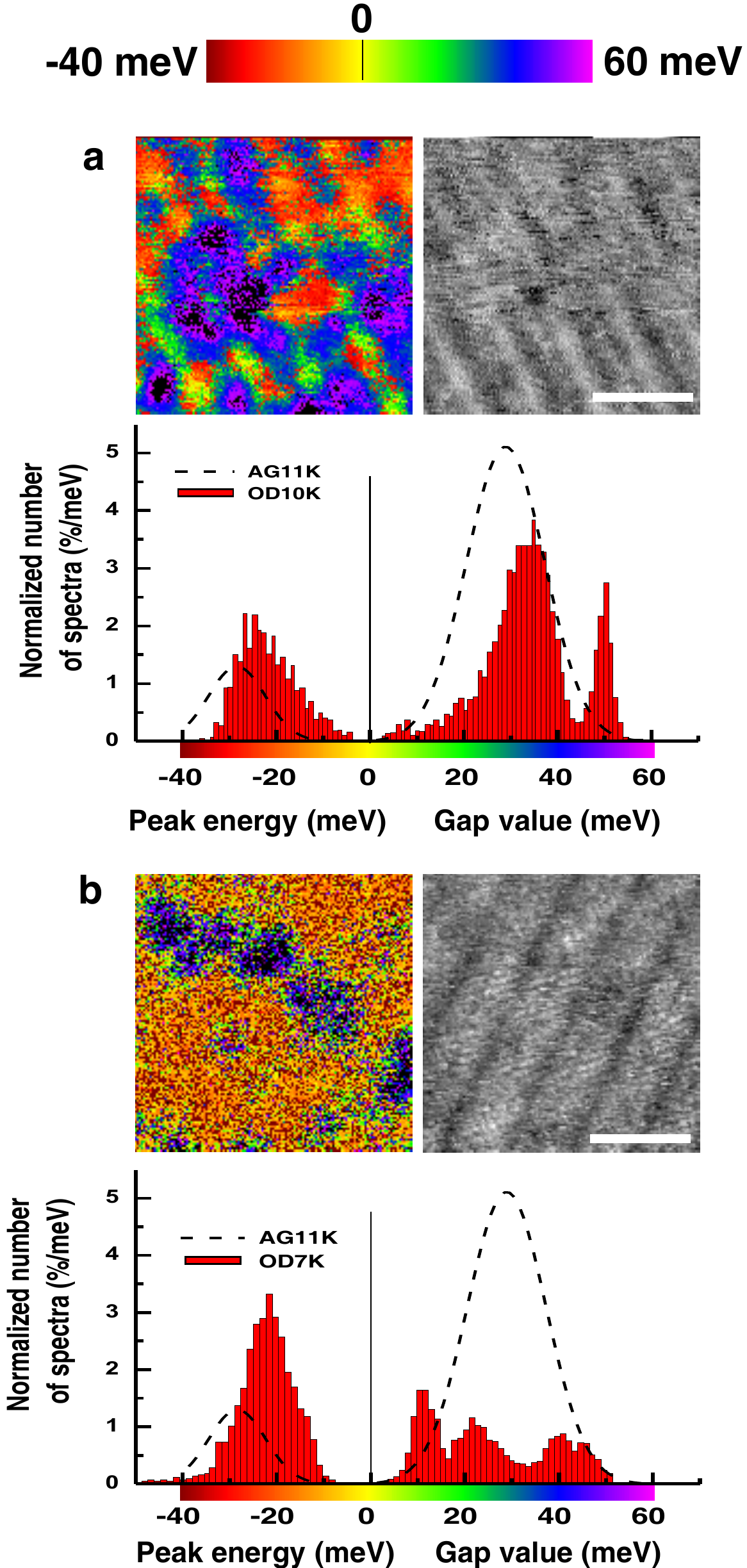}
\caption{\label{fig:fig2}
\textbf{Evolution of the Bi-2201 spectroscopic properties with overdoping.} The top-left panels show the $E_p$-maps ($E_p$ is the peak position for single-peak spectra, and half the peak-to-peak distance for two-peak spectra) for \textbf{a}, an overdoped sample with $T_c=10$~K and $\Delta T_c=2.5$~K (OD10K), and for \textbf{b}, an overdoped sample with $T_c=7$~K and $\Delta T_c=1.5$~K (OD7K). The white bars correspond to 5~nm. The distributions of $E_p$ values presented in the histograms were normalized and compared with the histograms obtained in the least overdoped AG11K sample (dashed lines). The top-right panels are simultaneously acquired topographies: white regions are physically higher than black ones.  They show the superstructure common to all Bi-based cuprates \cite{Gao-1988, Giannini-2008}. Especially in OD10K, the smallest gaps (green) are preferentially located on the minima of the superstructure. A similar correlation between gap size and the superstructure has been reported in Bi-2212 \cite{Slezak-2008} and Bi-2223 \cite{Jenkins-2009}. Our measurements further show that the preferred location of single-peak spectra (red-yellow) is also the minimum of the superstructure, suggesting that the local doping level itself is modulated.
}
\end{figure}

Low-temperature spatially-resolved STS of an as-grown sample with $T_c=11$~K and a transition width $\Delta T_c=4.5$~K are depicted in Fig.~\ref{fig:fig1}. Most of the surface presents spectra with a clear superconducting gap in the 15--45~meV range. However, extended regions covering 11\% of the sample exhibit qualitatively different spectra, without a gap but with a single sharp peak 10--40~meV below the Fermi energy. A broadly peaked background has been seen earlier in Bi-2201 but only at high temperature \cite{Kugler-2001}. The observation of this new type of spectra at low temperature, which we attribute to the VHS, is our main finding. To each spectrum we assign a characteristic energy $E_p$ which is either the peak to peak gap value (if two peaks are detected) or the energy position of the single peak. The transition between the two characteristic kinds of spectra is remarkably gradual and reproducible, both spatially (Fig.~\ref{fig:fig1}d) and when the spectra are sorted according to the value of $E_p$ (Fig.~\ref{fig:fig1}a). The general trend is that as the height of the peak decreases, a coherence peak emerges at positive energy signaling the opening of the gap, and the negative-bias peak shifts to larger negative energy. The nanometer-sized red to yellow areas corresponding in Fig.~\ref{fig:fig1}b to single-peak spectra are all surrounded by green regions which have small gaps.

We have then acquired $E_p$-maps and the corresponding histograms for two overdoped samples (OD10K and OD7K; Fig.~\ref{fig:fig2}). Comparing the histograms reveal a steady increase in the number of single-peak spectra with doping, from 11\% at $T_c=11$~K to 28\% at $T_c=10$~K and 48\% at $T_c=7$~K. The mean peak position also shifts towards the Fermi energy with values $-28$, $-26$, and $-21$~meV at $T_c=11$, 10, and 7~K, respectively, as expected in a rigid-band picture. Along with the increased proportion of single-peak spectra with increasing overdoping, we also observe a decreased proportion of wide-gap spectra that are excluded from the histograms, as one or both coherence peaks are missing. These spectra (depicted as black pixels in the $E_p$-maps) have a proportion of  28\% in AG11K, 6\% in OD10K, and 5\% in OD7K. Much of this decrease is an effect of the post-annealing treatments which favor a more homogeneous distribution of oxygen, as confirmed by the drop of the superconducting transition width $\Delta T_c$.

\begin{figure}[t]
\includegraphics[width=0.8\columnwidth]{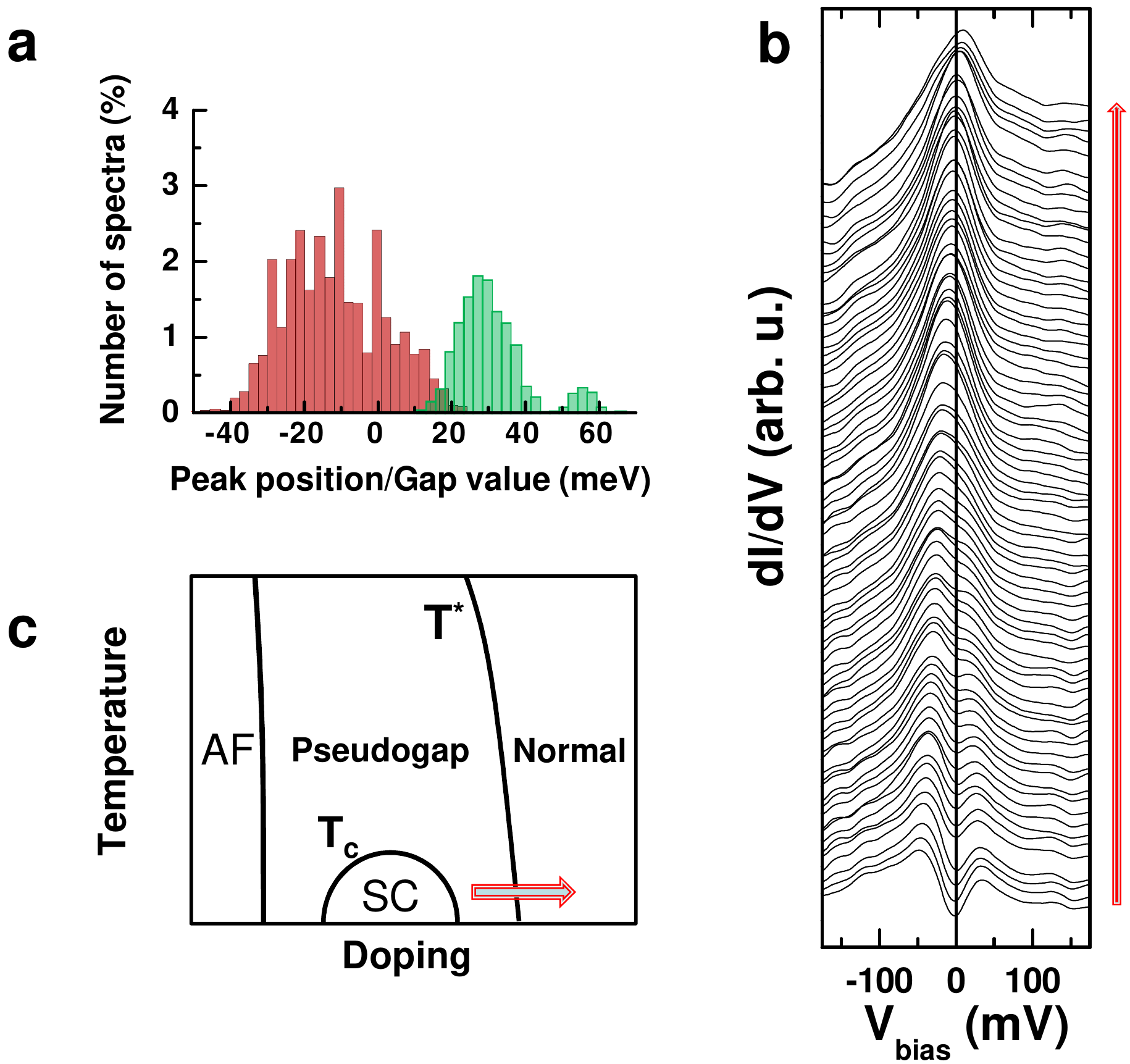}
\caption{\label{fig:fig3}
\textbf{\boldmath Spectral properties of strongly overdoped Bi-2201 ($T_c<1.8$~K) measured at 5~K.} (\textbf{a}) Distribution of $E_p$ values for single-peak spectra (red part of the histogram corresponds to the position of the single peak) and gapped spectra (green part of the histogram). (\textbf{b}) Representative spectra showing single peak at positive energy (top), single peak at negative energy, and gapped structures. This series was obtained by ordering the spectra according to the position of their maximum. (\textbf{c}) Interpretation of \textbf{b} as a transition from the pseudogap to the normal state. $T^*$, the frontier between normal and pseudogap states.
}
\end{figure}

Finally we have studied heavily overdoped sample in which $T_c$ has been reduced to less than 1.8~K. The proportion of single-peak spectra increases to $\sim 65$\% and some of these spectra even exhibit a peak at energy above $E_{\text{F}}$ (Fig.~\ref{fig:fig3}a). This suggests that a transition from a $(\pi,\pi)$-centered hole-like to a $(0,0)$-centered electron-like Fermi-surface takes place in the overdoped region near the border of the superconducting dome. Fits to ARPES data in cation-doped Bi-2201 have led to the same conclusion \cite{Kondo-2005}. As seen in Fig.~\ref{fig:fig3}b, as the gap vanishes and the peak moves to positive energy, the spectra develop a remarkable electron-hole asymmetry with a drastic suppression of spectral weight for occupied states. This trend mirrors the common phenomenology observed in all cuprates with increasing gap, namely an excess of spectral weight at negative energy. The data of Fig.~\ref{fig:fig3} were collected well above the bulk $T_c$ of the sample. It is therefore natural to interpret the series of spectra in Fig.~\ref{fig:fig3}b as a transition from the non-superconducting pseudogapped state to the un-pseudogapped normal state at fixed temperature (Fig.~\ref{fig:fig3}c). Since it is not possible to overdope Bi-2212 or Bi-2223 so as to reach a $T_c$ as low as in pure Bi-2201 \cite{Piriou-2008, Piriou-2010}, such a transition has not been observed before by STM and can now be investigated because of the identification of the VHS.

\begin{figure}[t]
\includegraphics[width=0.9\columnwidth]{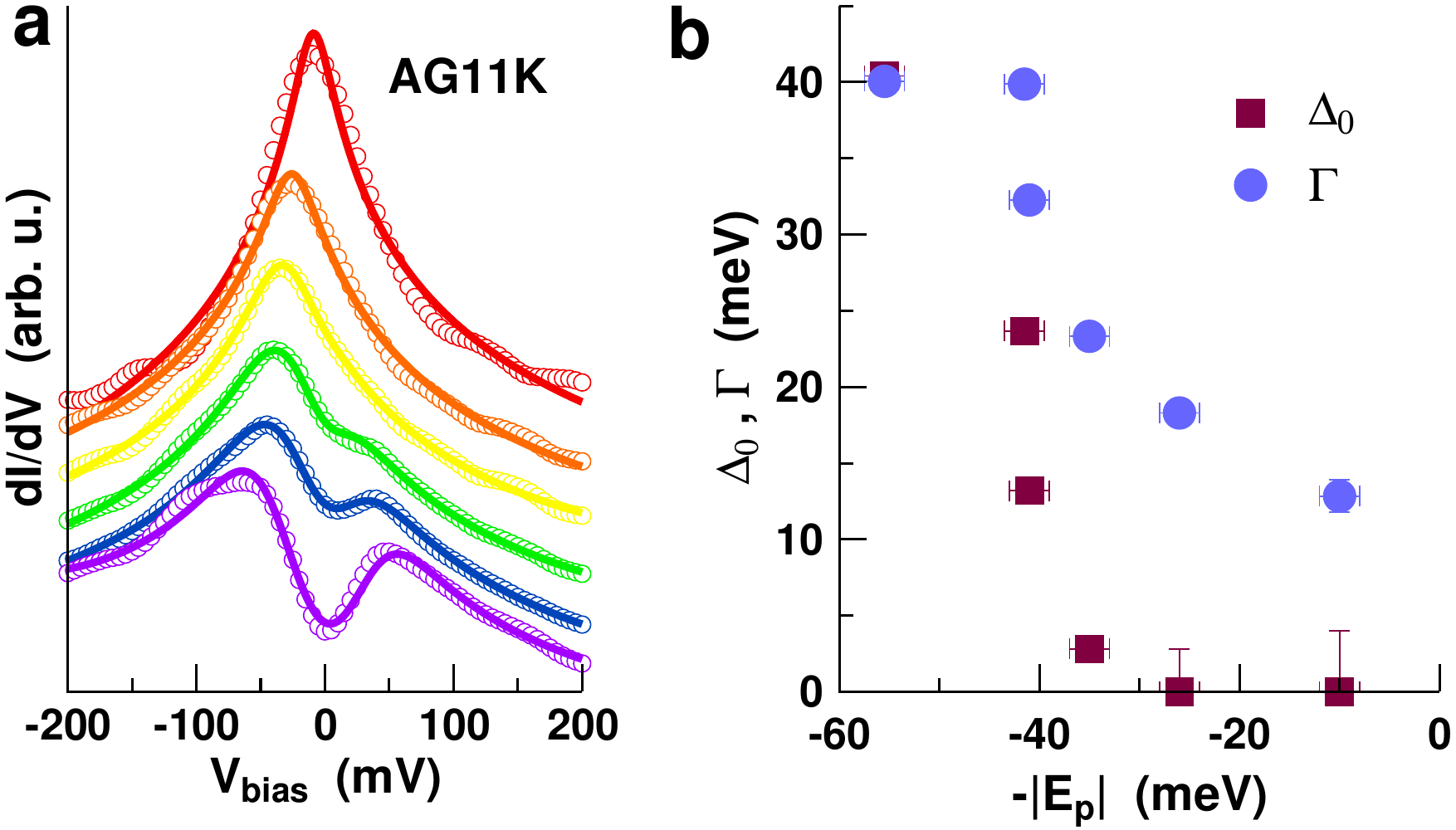}
\caption{\label{fig:fig4}
\textbf{Theoretical interpretation of Bi-2201 tunneling spectra.} (\textbf{a}) Comparison of representative $dI/dV$ curves in the as-grown sample (circles) with fits based on a simple model (lines). The model involves a two-dimensional tight-binding dispersion with a VHS, a $d$-wave BCS gap, and a constant scattering rate. The curves are offset vertically for clarity. (\textbf{b}) Squares: Amplitude of the BCS gap determined by the fits, as a function of $-|E_p|$. When $\Delta_0=0$ (single-peak spectra) the model corresponds to independent electrons subject to a constant scattering rate and the peak is the Van Hove singularity. Circles:  Trend of the phenomenological scattering rate $\Gamma$ as a function of $-|E_p|$.
}
\end{figure}

\subparagraph*{\hspace*{-\parindent}Theoretical modeling of the tunneling spectra.}

In the first STS studies of Bi-2201, the shape of the normal-state spectra was not addressed \cite{Kugler-2001, Boyer-2007}. More recently, sharp features in the tunneling characteristics have been reported in lead-doped Bi-2201 \cite{Chatterjee-2008}. The observed structure was a narrow peak very close to the Fermi level ($\pm2$~meV) superimposed on the superconducting gap. Such spectra were found at isolated locations on the surface and, based on earlier works in Bi-2212 \cite{Pan-2000a, Hudson-2001}, they were convincingly explained as native impurity resonances. The single-peak spectra in our measurements show no trace of a superimposed superconducting gap, cover extended regions of the surface, and have their maxima at a doping-dependent energy up to 40~meV below the Fermi level. In order to support our interpretation that the single-peak spectra are direct manifestations of the VHS in the DOS, we have modeled the $dI/dV$ curves obtained in our samples. Following previous analysis for Bi-2212 \cite{Hoogenboom-2003b} and Bi-2223 \cite{Levy-2008, Jenkins-2009}, we consider here the minimal tight-binding model featuring a VHS and a $d$-wave BCS gap. We then perform least-square fits to determine the band parameters, the gap size $\Delta_0$, and a scattering rate $\Gamma$. Fits to some representative spectra obtained on the as-grown sample are shown in Fig.~\ref{fig:fig4}. We observe that the model is able to follow the variety of measured spectral shapes with good accuracy. The amplitude of the BCS gap deduced from the fits is reported in Fig.~\ref{fig:fig4}b. The fits correctly yield a vanishing gap for single-peak spectra. We thus find that the single-peak spectra exhibit the same shape as the VHS in a two-dimensional tight-binding band. In general, the spectra measured in Bi-2201 are considerably broader than in Bi-2212 or Bi-2223. This hallmark can be captured by a phenomenological energy-independent scattering rate $\Gamma$.  The typical values of $\Gamma$ are indeed larger than in other Bi-based cuprates where $\Gamma\lesssim10$~meV \cite{Hoogenboom-2003b, Levy-2008, Jenkins-2009}, and we observe a tendency of $\Gamma$ to follow the trend of the gap, as expected if both would result from the same pairing interaction.

\section*{Discussion}

In Bi-2212 and Bi-2223, the bi- and tri-layer compounds of the same family, the spatial variations of the tunneling spectra have been ascribed to inhomogeneities of the doping level on a nanometer scale \cite{Fischer-2007}. In this scenario the single-peak structure we observe would correspond to extremely overdoped regions which are non-superconducting at 5~K. One thus expects that the proportion of such spectra increases in samples with higher doping levels. We have confirmed this expectation by studying overdoped samples with various doping levels. The STM observation of the VHS in a low-$T_c$ cuprate superconductor provides a clear-cut experimental answer to a long-standing theoretical question. Our results confirm that the STM tunneling conductance is proportional to the \textit{full} electron DOS, not only to the ``superconducting DOS'', and disqualify qualitative arguments used to claim the opposite. This must be taken seriously for the interpretation and modeling of STM tunneling spectra in materials with singularities in the electron dispersion.

The appearance of the VHS as a sharp peak in Bi-2201 poses the question of its signature in the spectra of other Bi-based cuprates. It has been shown in Refs~\onlinecite{Eschrig-2000, Hoogenboom-2003b, Levy-2008} that the coupling to the $(\pi,\pi)$ spin resonance suppresses the VHS peak and gives rise to the strong dip in Bi-2212 and Bi-2223 spectra. As suggested by ARPES \cite{Wei-2008} and our measurements, if a dip is present in pure Bi-2201, it must be very weak (see Fig.~\ref{fig:fig4}a). Within a spin-fluctuation pairing scenario, we may tentatively ascribe the low $T_c$ of pure Bi-2201 and the absence of dip in the DOS---hence the presence of the VHS---to a weak coupling to spin fluctuations. Interestingly in cation-doped Bi-2201, which exhibit higher critical temperatures, the VHS has not been seen \cite{Sugimoto-2006, Wise-2008, Wise-2009}.

Our observation in the heavily overdoped sample (Fig.~\ref{fig:fig3}) confirms the presence of the pseudogap phase well into the overdoped region in Bi-2201 \cite{Kugler-2001}. In a concurrent version of the phase diagram without a pseudogap in the overdoped region, the gapped spectra in Fig.~\ref{fig:fig3}b would have to be interpreted as regions that are superconducting at $T=5~\text{K}>T_c$ because of inhomogeneities. In this scenario, a mechanism different from the pseudogap would be needed to explain the anomalous large gap of $\sim 30$~meV, 30 times larger than the expected value for a BCS superconductor with a $T_c$ of 5~K. Furthermore  the observation above $T_c$ of the coexistence of gapped and un-gapped regions strongly supports the existence of the pseudogap in the overdoped side and questions the universality of a cuprate phase diagram in which the pseudogap phase would cross the superconducting dome.

\section*{Methods}

\subparagraph*{\hspace*{-\parindent}STS measurements.}

STS measurements were performed in a home-built STM placed in a low-temperature and ultra-high-vacuum environment similar to the one reported in Ref.~\onlinecite{Kugler-2000a}. The crystals were cleaved at room temperature in a $\sim10^{-9}$~mbar atmosphere, and immediately cooled down to 5~K. The bias voltage was applied to the sample, while the ground electrode was a mechanically-sharpened iridium tip. The tunneling current was 0.25~nA and the regulation bias was 0.4~V. Conductance maps were obtained by numerical differentiation of the $I(V)$ curves.

\subparagraph*{\hspace*{-\parindent}Modeling of the spectra.}

The $dI/dV$ curves were modeled as the thermally broadened DOS, $dI/dV\propto\int d\omega\,[-f'(\omega-eV)]N(\omega)$, where $f$ is the Fermi function and $N(\omega)$ is the BCS DOS given by $N(\omega)=\sum_{\vec{k}}(-1/\pi)\text{Im}\big[\omega-\xi_{\vec{k}} +i\Gamma-|\Delta_{\vec{k}}|^2/(\omega+\xi_{\vec{k}}+i\Gamma)\big]^{-1}$. The $d$-wave gap reads  $\Delta_{\vec{k}}=\Delta_0(\cos k_xa-\cos k_ya)/2$, and for the dispersion we used the five-neighbor tight-binding expression $\xi_{\vec{k}}=2t_1(\cos k_xa+\cos k_ya)+4t_2\cos k_xa\cos k_ya+2t_3(\cos 2k_xa+\cos 2k_ya)+4t_4(\cos 2k_xa\cos k_ya+\cos k_xa\cos 2k_ya)+4t_5\cos 2k_xa\cos2k_ya-\mu$ with $a$ being the lattice parameter. The $\vec{k}$-mesh involves $1024\times1024$ points. The model parameters were determined by least-square fits to the STM spectra. Given the restricted energy range of the experimental data, the fits can not reliably determine the overall bandwidth controlled by $t_1$. We therefore kept $t_1$ fixed to the ARPES value \cite{Kondo-2005} $t_1=-213$~meV for all spectra.

\section*{Acknowledgements}

This work was supported by the Swiss National Science Foundation through Division II and MaNEP. The authors thank Y. Fasano for fruitful discussions.

\bibliography{.}

\end{document}